# How to improve the quality of comparisons using external control cohorts in single-arm clinical trials?

**Auteurs**

Jérôme Lambert (1,2), Etienne Lengline (3), Raphaël Porcher (4), Rodolphe Thiébaut (5,6), Sarah Zohar (7,8,9), Sylvie Chevret (1,2)

**Institutions & Affiliations:**

1 Biostatistical Department, Hôpital Saint-Louis, AP-HP, Paris, France
2 INSERM, ECSTRRA Team, UMR1153, Univeristé Paris Cité, Paris, France
3 Hematology Department, Hôpital Saint-Louis, AP-HP, Paris, France
4 Center for Clinical Epidemiology, Hôtel-Dieu, AP-HP, Paris, France
5 Medical Information Department, CHU Bordeaux, France
6 Univ. Bordeaux, INSERM U1219, INRIA SISTM, Bordeaux, France
7 INSERM, Centre de Recherche des Cordeliers, Université Paris Cité, Sorbonne Université , Paris, France
8 Iniria, HeKA, Inria Paris, France
9 Université Paris Cité, INSERM, INRAE CRESS-UMR1153, Paris, France

**Acknowledments of research support**

**Corresponding author:** Chevret S, sbim- Saint Louis hospital, 1 ave Claude Velelfaux 75010 Paris, France, 33 1 42 49 97 42, sylvie.chevret@u-paris.fr

**Running head**. How to use external controls for single-arm trials

**Word count  3,915 (<4000)**

**Abstract**     275 words  (<275)

**PURPOSE**

Providing rapid answers and early acces to patients to innovative treatments without randomized clinical trial (RCT) is growing, with benefit estimated from single-arm trials. This has become common in oncology, impacting the approval pathway of health technology assessment agencies. We aimed to provide some guidance for indirect comparison to external controls to improve the level of evidence following such uncontrolled designs.

**METHODS**

We used the illustrative example of blinatumomab, a bispecific antibody for the treatment of B-cell ALL in complete remission (CR) with persistent minimal residual disease (MRD). Its approval relied on a single-arm trial conducted in 86 adults with B-cell ALL in CR, with undetectable MRD after one cycle as the main endpoint. To maximize the validity of indirect comparisons, a 3-step process for incorporating external control data to such single-arm trial data is proposed and detailed, with emphasis on the example.

**RESULTS**

The first step includes the definition of estimand, i.e. the treatment effect reflecting the clinical question. The second step relies on the adequate selection of external controls, from previous RCT or real-world data (RWD) obtained from patient cohort, registries, or electronic patient files. The third step consists in chosing the statistical approach targeting the treatment effect of interest, either in the whole population or restricted to the single-arm trial or the external controls, and depending on the available individual-level or aggregrated external data.

**CONCLUSION**

Validity of treatment effect derived from indirect comparisons heavily depends on carefull methodological considerations that are included in the proposed 3-step procedure. Because the level of evidence of a well conducted RCT cannot be guaranteed, post-market authorization evaluation is even more important than in standard settings.

**Introduction**

Thanks to the implementation of accelerated approval mechanisms by regulatory agencies such as Food and Drug Administration (FDA) breakthrough therapy designation and European Medicines Agency (EMA) accelerated assessment, uncontrolled trials are increasingly used by manufacturers as the designs of choice for new treatment evaluation seeking marketing authorization. Such accelerated processes, with no attempts to go up to any randomized trial, have become common in some therapeutic areas such as oncology, with biomarkers based cancer therapies and targeted antibodies, and impact the approval pathway,[1] based on weak and limited evidence.[2, 3] This was notably exemplified with immune checkpoints inhibitors, where nine of the 10 accelerated approvals involved single-arm trials with response rate as the main endpoint.[1, 4]

Accelerating clinical research with fewer patients involved and reduced costs, by contrast to the standard and separated phase I-II-III trials, could appear justified both from the patient and public health view.[5] Indeed, in some situations where there is an unmet medical need, a randomized clinical trial (RCT) may be difficult for a patient and clinician to accept, as long as preliminary evidence of effect may have been obtained in the earlier phases of development. Abandoning RCTs often relies on rejecting the 'clinical equipoise', that is the uncertainty of the scientific community around the benefit of the new treatment over the standard,[6] with two main arguments: 1/ the absence of an active identified or universally accepted comparator ["*This would have required including different heterogeneous treatments in the control arm ...*"] 2/ the implicit consideration that the new treatment may allow a paradigm shift in the goal of care ["*X treatment is curative, unlike existing treatments, used as salvage treatment and with suspensive effect*"] and brings more benefits than armfulness.

However, beside situations of quasi-deterministic disease evolution where nearly 0 or 100% of patients respond, relying on the observed "before-after" patient status to define treatment effect is well known to be biased, ignoring the Hawthorne, regression to the mean, placebo effects, and selection biases.[7] Moreover, the effect size of new molecules is mostly small, based on poorly relevant outcomes,[8] that may introduce additional classification biases due to the open nature of the design.

Thus, it is mandatory to account for such uncertainties, not to compromise scientific evidentiary standards of evidence-based medicine.[9] To increase the level of evidence in these early settings for regulatory and Health Technology Assessment (HTA) agencies that have to face early decision-making for drug approvals, the use of external comparisons, i.e. looking for external controls should be promoted.[10] However, as recently reported in oncology trials,[11] to maximize its validity, such comparisons require a careful implementation of innovative statistical methods accounting for between-group variation and selection biases. Appropriate methodologies depend on the availability and type of external data, either individual-level data or only aggregated data.[12] However, while many authors warned against the misuse of each approach and methodological issues from the use of external controls,[13–16] none has detailed the whole process, including the underlying assumptions for leveraging those data.

To improve the level of evidence of treatment effect following uncontrolled designs, we aimed to detail the 3-step process of incorporating external control data to single-arm trial data (Figure 1). First, the specifications of key attributes or "estimands", in line with the objectives,[17] should be defined according to the principles of such "emulated" target trials.[18] Second, selection of the controls should consider the various sources of external controls to adequately mimic the lacking randomized experiment. Specific statistical considerations arise, according to data type, i.e. individual or agregated patient data. The last step consists in the indirect comparison itself, based on different methods according to the available data and the targeted treatment effect. A motivating example is used to illustrate the 3-step process.

**Motivating example**

On March 29, 2018, the FDA granted accelerated approval to blinatumomab (Blincyto, Amgen Inc.), a bispecific T-cell engager antibody targeting CD19, an antigen expressed on B-lineage acute lymphoblastic leukemia (ALL) cells, for the treatment of B-cell precursor ALL in first or second complete remission (CR) with persistent or recurrent minimal residual disease (MRD) greater than or equal to 0.1% in adults and children. The approval only relied on the BLAST trial (NCT-01207388), an open-label, single-arm, phase 2 trial, conducted in 86 adults with B-cell precursor ALL in hematologic CR.[19] Efficacy was based on achievement of undetectable MRD within one cycle of blinatumomab and hematological relapse-free survival (RFS). Overall, undetectable MRD was achieved by 52 (85%) patients in first CR (95% CI, 74–93%) and 18 (72%) patients in second CR (95%CI, 51–88%). The median estimated hematological RFS was 35.2 months for patients in first CR and 12.3 months for patients in second CR.

We explored the possibility of performing external comparisons based on relevant pertinent comparators, notably a straightforward up-front allogeneic hematopoietic stem-cell transplantation (HSCT).

**Step 1- Definition of Estimands**

According to the International Conference for Harmonization (ICH), an estimand is a precise description of the treatment effect reflecting the clinical question, that should inform design and analysis under five attributes: target population, treatment, endpoint, intercurrent events and population level summary of the treatment effect measured against some valid comparator. First described for RCT,[20] its principles can be easily extended to observational studies.[21, 22]

When there is no marked doubt about the comparability of the treated and control populations, they can be assumed similar. Then, down-weighting the external control data allows handling the decreased level of evidence from the external source, either using power prior models,[23–25] or metaanalytic approaches,[14] from Bayesian inference or weighted statistics.

However, most of the time, populations differ in characteristics that may also affect the outcome, which are termed "confounders" (Box 1). Ignoring those difference results in misleading inference because of confounding bias.[26] Indeed, any differences observed in outcomes could no longer be attributed to differences in treatments but in confounders.

Thus, reaching balance in confounders is at the core of causal inference in observational studies, where adapted statistical methods have been widely developed since the early 80s. Those methods allow handling multiple confounders through the propensity score (PS), i.e., the probability of being in the treatment group, conditional on the set of observed confounders.[27] Then, samples are matched or weighted to minimize the discrepancy in observed confounders between treatment groups: in other words, individuals are given individual "balancing" weights,[21] derived from their PS, in order to under- or over-represent the characteristics of their treatment group compared to the other (Figure 2). Under different assumptions of conditional independence, consistency, and common support (Box 2), valid estimators can be directly derived from weighted data.

In the particular setting of the comparison of single-arm vs external control groups, those methods could be directly used. However, the target population should first be defined, that impacts the definition of weights and the targeted treatment effect (Table 1). Indeed, one may focus on the average treatment effect (ATE) in the population represented by the combined single-arm and external control groups, that would be observed by switching every unit in the whole population from one treatment to the other, or the average treatment effect in the treated (ATT), obtained by only switching the treated to the control group, or the average treatment effect in the control (ATC) conversely.

In evaluating the benefit of blinatumomab versus HSCT as recommended in patients with ALL in CR at the time,[28] the ATE, corresponding to switching every unit in the study population from blinatumomab to HSCT and reciprocally, may result in the effect of an infeasible intervention. By contrast, chosing the ATT targets the treated population, that is, those included in the single-arm trial, attempting to answer to 'what would have been the undetectable MRD of the patients treated by blinatumomab if they all have received a HSCT?'. The ATC provides the alternate answer to 'what should have been the undetectable MRD of the patients from the HSCT group had they received blinatumomab instead?'. Unfortunalely, such targets are rarely reported and justified.

**Step 2- Selection of External control data**

Once the objectives, the target population and the target treatment effects over some appropriate comparator have all been defined and justified, one may look for external, sometimes called "synthetic",[29] controls.

In line with the objective, the closeness of the external population to the targeted population should be first required to avoid the risk of substantial biases. It could be evaluated using the acceptability criteria proposed by Pocock,[30] in terms of patient eligibility, methods for treatment evaluation, distribution of patient characteristics, and usual care. Selection of external controls should mimic the selection process of meta-analyses, with predefined eligibility criteria for including studies to insure patient similarity, relevant endpoints, and pertinent comparator.

External controls could be directly selected from pertinent and efficacious active arms from previously completed RCTs ,[12] or reconstituted from real-world data (RWD) such as Electronic Health Records (EHR), registeries or insurance claim datasets.

When external controls are selected from RCTs, it is likely that the potential comparator has been sponsored by another firm, so that only aggregated data are available. Pooled data from previous RCTs could also be used as external controls, as exemplified by the FDA that has approved a synthetic control generated from more than 22,000 previous studies, to be used in a Phase III glioblastoma cancer trial.[31] However, the control of confounders is more challenging with aggregated data.

When no available controls from previous trials are available, controls can be selected from RWD including observational cohorts, registries or EHR,[32] but also claims and prescription data.[33] Adequate period of time and centers, to control for potential time and center effects, should be first considered.[34] The closeness of populations is of particular concern in such an observational setting where there is an obvious "confounding-by-indication" bias due to the choice of treatment based on patient disease status. Of note, there has been evidence that participants to clinical trials are generally more selected, even within eligibility criteria, treated in more experienced or specialized centers, and with a better follow-up than in routine care settings.[35] In many chronic diseases, there is also no obvious single timepoint for treatment decision, and patients could be treated following a number of previous decisions of no treatment, or use of other treatments after failure.[34] When population differ in terms of time of treatment decision in the course of the disease, "immortal time bias" or "time-lag bias" could be additionally introduced.[36, 37] For instance, CAR-T cells administered in a single dose after a variable period of time, between patient selection, leukapheresis of the patient cells and a possible bridge chemotherapy, raise concerns about the comparison of cohorts with different in start dates of follow-up.[38] Once RWD sources of control data are found, their validity should be measured through the assessment of risk of bias, often lacking. As reported recently, based on publicly available FDA reviews of medical products, most reasons why RWD did not contribute to regulatory decision-making relied on lack of pre-specification of study design and analysis as well as data reliability and relevancy concerns.[39]

In the blinatumomab example, a comparison to external controls obtained from a retrospective cohort study of adult patients with Ph-negative BCP ALL in hematological CR with MRD persistence who received standard-of-care treatment at various European study groups facilities, was actually added to the FDA application. Key eligibility criteria of the BLAST trial were applied to define the eligible subset among the observational study participants. Only patients in first CR (both for BLAST participants and external controls) were considered, which was noted as a limitation by FDA reviewers, the target population being different from the original intended population. Other limitations noted were the non-contemporaneous controls (the observational study having started 10 years before the trial), differences in proportions of patients receiving HSCT, in measurement of the outcome (locally or centrally performed, with different techniques, at distinct time points).

**Step 3- Methods for indirect comparison of single-arm and external control arm**

Last, after pertinent external data have been identified and quality checked, indirect comparaison of the single-arm trial and the external control should be done using appropriate statistical methods, checking underlying assumptions. Methods mostly depend on whether the control data have been measured at the individual-level or aggregated level.

Individual-level external control data

The availability of individual-levels data for both groups allows estimating the PS to balance the confounders of the treated (trial) group and the (external) control group using weighting or matching (Table 1). When external individual-level data are obtained from observational data, additional weights may be used to incorporate the decreased level of evidence of the controls as compared to the trial data.[41]

The most common approach to estimating inverse probability of treatment weights (IPW) is to estimate the PS through logistic regression, though approaches such as machine-learning, could be alternatively used,[42] and then take its reciprocal. Such weights target the ATE of the underlying population (Figure 2). Unfortunately, the sample is often a "convenience" sample and does not represent any population of scientific interest, by contrast to surveys from which such methods have been derived.

To focus on the treated population and estimating the ATT, treated patients are given a weight of 1 while control patients are weighted by the odds of being treated. Such ATT weights are sometimes referred to as "standardized mortality ratio weights".[43]

For both types of weights, the challenge of extreme propensities has been identified as a primary downside of weighting, with no clear definition of the resulting ambiguous target population.[44] Indeed, due to the assumption of common support, it should be found at least one control and one treated at each value of the PS so that the distributions overlap; otherwise, methods that address non-overlap, such as trimming or down-weighting data in regions of poor data support, excluding or censoring weights at some extreme percentiles, change the estimand so that inference cannot longer target the population of interest. Thus, balancing weights have been proposed as a simple way to define, based on specific tilting functions, individual weights and resulting target population,[45] integrating most approaches including the PS-matching,[27] widely used as a way of estimating the ATT. Besides IPW and ATT weights, "overlap weights" were recently proposed as a way of focusing on the population for which observed confounders have been adequately balanced (Table 1). As stated above, all weighted samples differ in terms of underlying target population, as illustrated in the observed patient characteristics (Figure 2).

For the blinatumomab example, an IPW-based analysis of external controls was actually added to the FDA. The eligible subset among the study participants were compared to the BLAST study participants using IPW, with a set of additional sensitivity analyses. The original IPW analysis presented to the FDA only included patients in CR1 (both for BLAST participants and external controls). Besides those exposed above, the main limitation was the potential risk of bias with respect to unmeasured and unknown confounders.

Aggregated external control data

When control data come from clinical trials not sponsored by the manufacturer own product of the single-arm trial, it is not rare that only published aggregate data are available. In this setting, only summary measures of both the confounders and outcomes are at most available. Of note, for time-to-event data, some sorts of individual-level data can be extracted from published Kaplan-Meier curves using digitization,[46] but individual-level data on confounders would still not be obtained.

To deal with such aggregated control data, unanchored population-adjusted indirect comparisons have been proposed, the two most popular methods being Matching-Adjusted Indirect Comparison (MAIC) [47] and Simulated Treatment Comparison (STC).[48]

MAIC is a reweighting method similar to IPW, though targeting the control population. The principle is to reweight the individual-level data such that the characteristics of the treated are balanced with those of the patients from the control data, with weights estimated from PS of being treated, that is, on the individual-level trial data. The resulting target population is that of the external dataset, thus estimating the ATC (Table 1). Of note, the PS cannot be estimated as usual, but using alternate method of moments or entropy balancing, both shown equivalent.[49]

In STC, individual-level data are used to model the relationship between predictors and outcome of the single-arm trial, then using the model to estimate outcomes in external controls. The relative effect of the control over the treated group is then estimated by comparing this expected outcome to the observed aggregated outcome of the control.

In these unanchored comparisons, coverage of 95% confidence interval of treatment effect estimate has been showed to be suboptimal without clear explanation, maybe in relation to assessing variability in weights estimation. Both MAIC and STC rely on the strong assumption of conditional constancy of absolute effect, i.e., that the absolute treatment effect is constant at any level of the effect modifiers and prognostic variables, and that all effect modifiers and prognostic variables have been observed, otherwise the estimates are biased.[50] This assumption is widely recognized as very hard to meet. This is a main issue when the mechanism of action of single-arm treatment is dependent on a biological target not involved in the response in the control cohort. Thus, providing information on the likely bias resulting from unobserved prognostic factors and effect modifiers distributed differently across the trials, is mandatory.

An important limitation is that MAIC or STC, as currently proposed, are only able to provide estimates in the target population represented by the external comparator population and not that of the single-arm trial of interest. For any other target populations, a supplementary assumption, the shared effect modifier, is needed.[50]

Such indirect comparisons require additional recommendations. First, evidence that absolute outcomes can be predicted with sufficient accuracy in relation to relative treatment effect should be provided. Moreover, the choice of the outcome scale is critical and should be justified, since effect modifier status is scale-specific.

For the blinatumomab example, external comparison could have been performed using published data on 272 patients with ALL in CR and quantifiable MRD from ALL study groups managed in national protocols 2000–2014.[51] Baseline characteristics on age, sex, time from diagnosis, year of diagnosis, WBC count and cytogenetics at diagnosis and MRD status, could have been used as potential confounders. Estimated treatment benefit on RF and OS could have been considered.

**Discussion and Perspectives**

Standard strategy of treatment evaluation is based on successive phases, from phase I dose-findings dedicated to tolerance up to phase III RCT comparing the effects of the new treatment to a pertinent comparator on exchangeable groups of patients. Nevertheless, providing rapid answers in evaluating a new treatment outside this standard strategy is growing.[52] Nowadays, the use of single-arm clinical trials as the sole source of evidence provided by the pharmaceutical firms to obtain -at least tempororay- drug approvals, is accepted by regulatory agencies in some indications or populations. This may appear contradictory to the huge statistical literature reporting its many sources of bias since the early 80s.[53]

To improve the level of evidence and decrase the uncertainty of such uncontrolled trials, comparisons using external or "synthetic" controls have been reported, for instance in large B-cell lymphoma,[54] metastatic non-small cell lung cancer,[55] or glioblastoma.[56] Such comparisons require a careful and sometimes complex implementation to be valid, as reported in the recent review of methods for observational data.[57] In the specific setting of single-arm trials, we aimed to report how to enhance the evidence from such trials, by incorporating and leveraging external data as a "synthetic" control-arm, whichever data come from other trials, observational studies or the real-world, and available at the individual-level or not.

We provided some guidance in incorporating external controls by defining a 3-step process, with the aim to stop the sequence whenever some target or underlying assumption could not be satisfied. First, the targeted population, pertinent comparator and measure of treatment effect should be clearly delineated. Secondly, all sources of the target controls should be considered, with selection adequately performed with respect to the population, endpoint and treatment decision. Last, the method of analysis should be justified based on the type of available data, and on the underlying population. External controls entail merging different sources of data, which may complicate the verification of causal assumptions and adequate control for confounding.

In all cases, and given the risk that analyses would be data-driven and adapted ad-hoc, the statistical analysis plan for such an incorporation should be publicly issued before analysis, and only external controls recruited after that publication would be used in the comparisons, in a similar approach as registered reports.[58] The principled framework of emulating a target trial combining principles of clinical trials and causal methods to control for confounding appears particularly adequate in this situation.[18, 59]

We mostly considered methods derived from propensity score, though other approaches could be also considered. First, with the use of individual-level data, *g*-computation—or regression standardization— developing risk prediction models for the outcome under each treatment (experimental or control), then using individual predictions to estimate the ATE, ATT or ATC,[60–62] has been proposed. Otherwise, "double-robust" or "augmented IPW" estimators combine IPW and prediction models of *g*-computation.[63] To our knowledge, these approaches have not been used for regulatory approval with external controls, but remain valid promising alternatives.

Other issues such as time-dependent biases may exist.[36, 37] Statistical methods have been proposed to control for confounding in such a time-dependent situation, such as marginal structural models,[64] or risk-set matching.[65] However, these methods require a follow-up of the whole cohort including treated and control individuals, which is not the case when using external aggregated data for controls. How to adequately control for time-dependent biases with external controls has therefore also to be considered.

Otherwise, the desire of rapid answers may also raise the interest of new approaches for which both interventional and control groups may lack. *In silico* trials referred to models and simulations used for drug and other interventions development.[66, 67] Their general idea is to build a mechanistic model based on biological knowledge such as drug mechanism of action, and parameter values using available data or specific experimentations. Once defined, the model is used to explore new scenarios that could have been evaluated in clinical trials. This can lead to the choice of one or several strategies among many to be further evaluated. As an example, such an approach has been used for the development of an immune intervention in HIV-infected patients based on the interleukin-7 using an ODE-based model,[68] that has served to select and define next trials.[69] Although promising, these approaches are challenging because of the limitations due to model misspecifications and parameters identifiability.

In summary, when reporting results from a single-arm trial with the aim of marketing authorization, providing some external comparison to controls should be mandatory, adequately done and reported. It should be kept in mind that such indirect comparisons aim to mimic the missing randomized clinical trial. Thus, it includes, not only the report of indirect comparisons, but also the clear justification of the target population throughout eligibility criteria, of the potential sources of available data, of the method used for estimation, and of the checking of all the underlying causal assumptions. Only the respect of all these steps may provide a correct level of evidence although it could not be guaranteed that it will reach the level of a well conducted RCT. It is therefore at utmost importance to follow the effectiveness of any interventions after market authorization that has been evaluated through single-arm trial with propre post-market surveillance system. Finaly, our approach can be generalized in other indications and populations for which single-arm trials have lead to regulatory approval.

## References

**1**. Beaver JA, Pazdur R: "Dangling" Accelerated Approvals in Oncology. N Engl J Med 384:e68, 2021
**2**. Naci H, Davis C, Savović J, et al: Design characteristics, risk of bias, and reporting of randomised controlled trials supporting approvals of cancer drugs by European Medicines Agency, 2014-16: cross sectional analysis. BMJ l5221, 2019
**3**. Hatswell AJ, Freemantle N, Baio G: The Effects of Model Misspecification in Unanchored Matching-Adjusted Indirect Comparison: Results of a Simulation Study. Value Health 23:751–759, 2020
**4**. Beaver JA, Pazdur R: The Wild West of Checkpoint Inhibitor Development. N Engl J Med 386:1297–1301, 2022
**5**. Zelner J, Riou J, Etzioni R, et al: Accounting for uncertainty during a pandemic. Patterns N Y N 2:100310, 2021
**6**. Shamy M, Fedyk M: Why the ethical justification of randomized clinical trials is a scientific question. J Clin Epidemiol 97:126–132, 2018
**7**. Sedgwick P: Before and after study designs. BMJ 349:g5074–g5074, 2014
**8**. Ribeiro TB, Colunga-Lozano LE, Araujo APV, et al: Single-arm clinical trials that supported FDA accelerated approvals have modest effect sizes and at high risk of bias. J Clin Epidemiol S0895435622000257, 2022
**9**. Howick, J: The philosophy of evidence-based medicine. Chichester, Wiley-Blackwell, 2011
**10**. Davi R, Mahendraratnam N, Chatterjee A, et al: Informing single-arm clinical trials with external controls. Nat Rev Drug Discov 19:821–822, 2020
**11**. Collignon O, Schritz A, Spezia R, et al: Implementing Historical Controls in Oncology Trials. The Oncologist 26:e859–e862, 2021
**12**. Goring S, Taylor A, Müller K, et al: Characteristics of non-randomised studies using comparisons with external controls submitted for regulatory approval in the USA and Europe: a systematic review. BMJ Open 9:e024895, 2019
**13**. Burcu M, Dreyer NA, Franklin JM, et al: Real-world evidence to support regulatory decision-making for medicines: Considerations for external control arms. Pharmacoepidemiol

Drug Saf 29:1228–1235, 2020
**14**. Schmidli H, Häring DA, Thomas M, et al: Beyond Randomized Clinical Trials: Use of External Controls. Clin Pharmacol Ther 107:806–816, 2020
**15**. Wang C, Berlin JA, Gertz B, et al: Uncontrolled Extensions of Clinical Trials and the Use of External Controls—Scoping Opportunities and Methods. Clin Pharmacol Ther 111:187–199, 2022
**16**. Yap TA, Jacobs I, Baumfeld Andre E, et al: Application of Real-World Data to External Control Groups in Oncology Clinical Trial Drug Development. Front Oncol 11:695936, 2022
**17**. ICH Harmonised Guideline E9(R1): Estimands and Sensitivity Analysis in Clinical Trials [Internet], 2017Available from: https://database.ich.org/sites/default/files/E9-R1_Step4_Guideline_2019_1203.pdf
**18**. Hernán MA, Robins JM: Causal Inference Chapman&Hall. Boca Raton, FL, 2019
**19**. Gökbuget N, Dombret H, Bonifacio M, et al: Blinatumomab for minimal residual disease in adults with B-cell precursor acute lymphoblastic leukemia. Blood 131:1522–1531, 2018
**20**. Ratitch B, Goel N, Mallinckrodt C, et al: Defining Efficacy Estimands in Clinical Trials: Examples Illustrating ICH E9(R1) Guidelines. Ther Innov Regul Sci 54:370–384, 2020
**21**. Li H, Wang C, Chen W, et al: Estimands in observational studies: Some considerations beyond ICH E9 ( R1 ). Pharm Stat pst.2196, 2022
**22**. Goetghebeur E, le Cessie S, De Stavola B, et al: Formulating causal questions and principled statistical answers. Stat Med 39:4922–4948, 2020
**23**. Hobbs BP, Carlin BP, Mandrekar SJ, et al: Hierarchical Commensurate and Power Prior Models for Adaptive Incorporation of Historical Information in Clinical Trials. Biometrics 67:1047–1056, 2011
**24**. Brard C, Hampson LV, Gaspar N, et al: Incorporating individual historical controls and aggregate treatment effect estimates into a Bayesian survival trial: a simulation study. BMC Med Res Methodol 19:85, 2019
**25**. Roychoudhury S, Neuenschwander B: Bayesian leveraging of historical control data for a clinical trial with time-to-event endpoint. Stat Med 39:984–995, 2020
**26**. Dron L, Golchi S, Hsu G, et al: Minimizing control group allocation in randomized trials using dynamic borrowing of external control data – An application to second line therapy for non-small cell lung cancer. Contemp Clin Trials Commun 16:100446, 2019
**27**. Rosenbaum PR, Rubin DB: The central role of the propensity score in observational studies for causal effects. Biometrika 70:41–55, 1983
**28**. Bassan R, Hoelzer D: Modern therapy of acute lymphoblastic leukemia. J Clin Oncol Off J Am Soc Clin Oncol 29:532–543, 2011
**29**. Seeger JD, Davis KJ, Iannacone MR, et al: Methods for external control groups for single arm trials or long-term uncontrolled extensions to randomized clinical trials. Pharmacoepidemiol Drug Saf 29:1382–1392, 2020
**30**. Pocock SJ: The combination of randomized and historical controls in clinical trials. J Chronic Dis 29:175–188, 1976
**31**. Spinner J: Medidata synthetic control arm lands FDA approval for cancer trial. [Internet], 2020Available from: https://www.outsourcing-pharma.com/Article/2020/11/19/Synthetic-control-arm-lands-FDA-approval-for-cancer-trial
**32**. Tan K, Bryan J, Segal B, et al: Emulating Control Arms for Cancer Clinical Trials Using External Cohorts Created From Electronic Health Record-Derived Real-World Data. Clin Pharmacol Ther 111:168–178, 2022
**33**. Cave A, Kurz X, Arlett P: Real-World Data for Regulatory Decision Making: Challenges and Possible Solutions for Europe. Clin Pharmacol Ther 106:36–39, 2019
**34**. Suissa S: Single-arm Trials with Historical Controls: Study Designs to Avoid Time-related Biases. Epidemiology 32:94–100, 2021

35. Rothwell PM: External validity of randomised controlled trials: "To whom do the results of this trial apply?" The Lancet 365:82–93, 2005
36. Suissa S: Immortal Time Bias in Pharmacoepidemiology. Am J Epidemiol 167:492–499, 2008
37. Suissa S, Azoulay L: Metformin and the risk of cancer: time-related biases in observational studies. Diabetes Care 35:2665–2673, 2012
38. Lin X, Lee S, Sharma P, et al: Summary of US Food and Drug Administration Chimeric Antigen Receptor T-Cell Biologics License Application Approvals From a Statistical Perspective. J Clin Oncol JCO.21.02558, 2022
39. Mahendraratnam N, Mercon K, Gill M, et al: Understanding Use of Real-World Data and Real-World Evidence to Support Regulatory Decisions on Medical Product Effectiveness. Clin Pharmacol Ther 111:150–154, 2022
40. Thorlund K, Dron L, Park JJ, et al: Synthetic and External Controls in Clinical Trials – A Primer for Researchers. Clin Epidemiol Volume 12:457–467, 2020
41. Bonander C, Humphreys D, Degli Esposti M: Synthetic Control Methods for the Evaluation of Single-Unit Interventions in Epidemiology: A Tutorial. Am J Epidemiol 190:2700–2711, 2021
42. Robbins MW, Davenport S: **microsynth** : Synthetic Control Methods for Disaggregated and Micro-Level Data in *R* [Internet]. J Stat Softw 97, 2021[cited 2022 Jun 7] Available from: http://www.jstatsoft.org/v97/i02/
43. Sato T, Matsuyama Y: Marginal Structural Models as a Tool for Standardization: Epidemiology 14:680–686, 2003
44. Crump RK, Hotz VJ, Imbens GW, et al: Dealing with limited overlap in estimation of average treatment effects. Biometrika 96:187–199, 2009
45. Li F, Thomas LE: Addressing Extreme Propensity Scores via the Overlap Weights [Internet]. Am J Epidemiol , 2018[cited 2022 Jun 7] Available from: https://academic.oup.com/aje/advance-article/doi/10.1093/aje/kwy201/5090958
46. Guyot P, Ades A, Ouwens MJ, et al: Enhanced secondary analysis of survival data: reconstructing the data from published Kaplan-Meier survival curves. BMC Med Res Methodol 12:9, 2012
47. Signorovitch JE, Wu EQ, Yu AP, et al: Comparative Effectiveness Without Head-to-Head Trials: A Method for Matching-Adjusted Indirect Comparisons Applied to Psoriasis Treatment with Adalimumab or Etanercept. PharmacoEconomics 28:935–945, 2010
48. Phillippo DM, Ades AE, Dias S, et al: Methods for Population-Adjusted Indirect Comparisons in Health Technology Appraisal. Med Decis Making 38:200–211, 2018
49. Phillippo DM, Dias S, Elsada A, et al: Population Adjustment Methods for Indirect Comparisons: A Review of National Institute for Health and Care Excellence Technology Appraisals. Int J Technol Assess Health Care 35:221–228, 2019
50. Phillippo DM, Ades T, Palmer S, et al: Methods for population-adjusted indirect comparisons in submissions to NICE. Decision Support Unit, ScHARR, University of Sheffield, NICE Decision Support Unit, 2016
51. Gökbuget N, Dombret H, Giebel S, et al: Minimal residual disease level predicts outcome in adults with Ph-negative B-precursor acute lymphoblastic leukemia. Hematology 24:337–348, 2019
52. Johnson JR, Ning Y-M, Farrell A, et al: Accelerated approval of oncology products: the food and drug administration experience. J Natl Cancer Inst 103:636–644, 2011
53. Spodick DH: The randomized controlled clinical trial. Am J Med 73:420–425, 1982
54. Banerjee R, Midha S, Kelkar AH, et al: Synthetic control arms in studies of multiple myeloma and diffuse large B-cell lymphoma. Br J Haematol 196:1274–1277, 2022
55. Menefee ME, Gong Y, Mishra-Kalyani PS, et al: Project Switch: Docetaxel as a potential

synthetic control in metastatic non-small cell lung cancer (mNSCLC) trials. J Clin Oncol 37:9105–9105, 2019

56. Sampson JH, Achrol A, Aghi MK, et al: MDNA55 survival in recurrent glioblastoma (rGBM) patients expressing the interleukin-4 receptor (IL4R) as compared to a matched synthetic control. J Clin Oncol 38:2513–2513, 2020

57. Xu R, Chen G, Connor M, et al: Novel Use of Patient-Specific Covariates From Oncology Studies in the Era of Biomedical Data Science: A Review of Latest Methodologies. J Clin Oncol JCO.21.01957, 2022

58. Naudet F, Siebert M, Boussageon R, et al: An open science pathway for drug marketing authorization-Registered drug approval. PLoS Med 18:e1003726, 2021

59. Hernán MA, Sauer BC, Hernández-Díaz S, et al: Specifying a target trial prevents immortal time bias and other self-inflicted injuries in observational analyses. J Clin Epidemiol 79:70–75, 2016

60. Hernan MA: Estimating causal effects from epidemiological data. J Epidemiol Community Health 60:578–586, 2006

61. Vansteelandt S, Keiding N: Invited Commentary: G-Computation-Lost in Translation? Am J Epidemiol 173:739–742, 2011

62. Snowden JM, Rose S, Mortimer KM: Implementation of G-Computation on a Simulated Data Set: Demonstration of a Causal Inference Technique. Am J Epidemiol 173:731–738, 2011

63. Bang H, Robins JM: Doubly robust estimation in missing data and causal inference models. Biometrics 61:962–973, 2005

64. Robins JM, Hernán MÁ, Brumback B: Marginal Structural Models and Causal Inference in Epidemiology: Epidemiology 11:550–560, 2000

65. Lu B: Propensity Score Matching with Time-Dependent Covariates. Biometrics 61:721–728, 2005

66. Pappalardo F, Russo G, Tshinanu FM, et al: In silico clinical trials: concepts and early adoptions. Brief Bioinform 20:1699–1708, 2019

67. Musuamba FT, Bursi R, Manolis E, et al: Verifying and Validating Quantitative Systems Pharmacology and In Silico Models in Drug Development: Current Needs, Gaps, and Challenges. CPT Pharmacomet Syst Pharmacol 9:195–197, 2020

68. Thiébaut R, Drylewicz J, Prague M, et al: Quantifying and Predicting the Effect of Exogenous Interleukin-7 on CD4+T Cells in HIV-1 Infection. PLoS Comput Biol 10:e1003630, 2014

69. Thiébaut R, Jarne A, Routy J-P, et al: Repeated Cycles of Recombinant Human Interleukin 7 in HIV-Infected Patients With Low CD4 T-Cell Reconstitution on Antiretroviral Therapy: Results of 2 Phase II Multicenter Studies. Clin Infect Dis Off Publ Infect Dis Soc Am 62:1178–1185, 2016
**Figure Legends**

**Figure 1**: Schematic 3-step process to be applied when incorporating external control data to single-arm trial data in order to maximize the validity of indirect comparisons

RCT: randomized clinical trial; RWD: real-world data; MAIC: Matched adjusted indirect comparison; STC: Simulated Treatment Comparison

**Figure 2**: Schematic representation of how data is weighted according to the estimand.

Suppose the original sample from the single-arm trial differs from the external controls in terms of patient severity, with 1 severe case (in green) over 4 in the trial compared to 3 over 4 in the external data. The objective is to modify the pooled data to obtain two groups where the proportion of severe cases is similar.

Most methods are based on the propensity-score PS, that is the probability of being in the trial for each patient. In this setting, each severe (green) patient is given a PS of 1/4 while each non-severe patient (blue) is given a PS of 3/4.

Inverse probability of treatment weight (IPW) consists in inversely weighting each individual in the original samples according to their probability of being in the original group, that is for the treated, the individual contribution of each patient is divided by their PS (thus resulting in adding 1/3 of fictive patient for each non-severe patient and 3 fictive individuals for severe cases, while in the external group, it is divided by 1 minus their PS, thus adding 1/3 of fictive patient for each severe patient and 3 fictive individuals for non-severe cases. This yields a weighted sample where the proportion of severe cases is similar in both groups (1/2), that differ from both original groups.

ATT weights consist in using all individuals from the single-arm trial (weight of 1), and weighting each individual in the external sample by the odds of being in the trial. This results in odds of (1/4)/(3/4)=1/3 in non-severe patients and of (3/4)/(1/4)=3 in severe cases, reaching a ¼ prevalence of severe cases in the pooled weighted dataset, that is that observed in the original treated patients from the trial.

ATC weights are conversely computed, with weights of 1 for each patients from the external data, while patients from the single-arm trial affected to a weight of (3/4)/(1/4) for severe trial patients, and (1/4)/(3/4) for non-severe patients. The resulting prevalence of severe cases is now that of the original external control group, that is, 3/4.

**Table 1:** Targeted population, weights, and estimands.

Let $e(x) = PS = Pr(T = 1|V)$, where T=1 for the single-arm treatment group, T=0 for the external control group, and V the set of observed confounders in both groups

| Method for controlling confounders | Weights (treated, untreated) | Targeted population | Tilting function | Estimand |
|---|---|---|---|---|
| Inverse weigthing | IPW= $(\frac{1}{e(x)}, \frac{1}{(1-e(x))})$ | Combined from the treated and untreated | 1 | ATE |
| | $(1, \frac{e(x)}{(1-e(x))})$ | Treated population | $e(x)$ | ATT |
| | $(\frac{1-e(x)}{e(x)}, 1)$ | Control population | $1- e(x)$ | ATC |
| | $(1-e(x), e(x))$ | Overlap population | $e(x)(1- e(x))$ | ATO |
| | $\frac{1\,(a<e(x)<1-a)}{e(x)}, \frac{1\,(a<e(x)<1-a)}{(1-e(x))}$ | Trimming population | $1\,(a<e(x)<1-a)$ | Non-specified |
| Matching | $\frac{\text{Min }(e(x),\,1-e(x))}{e(x)}, \frac{\text{Min }(e(x),\,1-e(x))}{(1-e(x))}$ | Matching population | Min $(e(x),\,1-e(x))$ | ATT |
| Matching adjusted indirect comparison | prop. $(\frac{1-e(x)}{e(x)}, 1)$ | Control population | $1- e(x)$ | ATC |

IPW: inverse probability of treatment weight; ATE : average treatment effect ; ATT: average treatment effect in the treated; ATC: average treatment effect in the control; ATO: average treatment effect in the overlap population

**Box 1: Glossairy**

- **Confounder :** Variables that affect both the treatment choixe and the outcome. Ignoring those variables in the comparison of treatment groups achives "confounding-by-indication" bias.

- **External control** is defined as group of patients external to the study, that differs from an internal control group consisting of patients from the same population assigned to a different treatment.

- **Immortal time bias** refers to the difference in estimation achieved by differences in selection time of the treatment groups, that favor the treated who "survived" up to the actual administration of the treatment. This can be controlled by the careful selection of the definition of time to selection in both groups, that should be as close as possible.

- **Propensity Score (PS):** The propensity score is the probability that a patient would receive the treatment of interest, based on the pre-therapeutic characteristics of the patient, the treating clinician, and the clinical environment. PS methods are used to reduce the bias in estimating treatment effects and allow investigators to reduce the likelihood of confounding when analyzing nonrandomized, observational data. Under several assumptions, such methods allow providing causal treatment effects.

- **Real World Evidence (RWE)** is the clinical evidence about the usage and potential benefits or risks of a medical product derived from analysis of RWD.

- **Real Worl Data (RWD)** are data relating to patient health status and/or the delivery of health care routinely collected from a variety of sources.

**Box 2: Causal assumptions**

- **Consistency** relates the observed outcome to potential outcomes that would be observed under each treatment compared, which forms the underlying statistical framework for the approach. Consistency is generally assumed as part of the causal model itself, but also implies that the treatments to be compared are well defined, and that there are no "hidden" versions of those, which may be arguable for external controls who may receive different treatments. In that case, consistency should be more considered as a distributional level, i.e., the distribution of the different versions of the "treatment" in the population.

- **No interference**, means that the effect of the treatment on the outcome of an individual is not affected by the other individuals being treated or not. It can be generally accepted for external controls, in particular because they are often selected in existing cohorts, registries or electronic health records, that would be unaffected by a limited-sample size trial being conducted, possibly in different locations or time period.

- **No unmeasured confounding**, means that the covariates measured for the trial participants and external controls comprise all those that are likely to affect the outcome and differ between both groups. This assumption is more challenging, since it requires in practice that all relevant prognostic factors would be recorded for both participants to the trial and external controls. Also, factors that may affect outcomes such as center-specific characteristics, socio-economic variables, environmental factors, standard of care, or health systems may not be available for either the trial participants or the external controls.

- **Positivity** or **common support**, means that all individuals have a non-null probability of receiving either treatment compared. External controls had virtually no chance of receiving the experimental treatment, but one should look at whether controls could have received the experimental treatment given their individual characteristics, had they been followed-up in an institution participating to the trial. This is not limited to trial's eligibility criteria, but one should also look at the other potential confounders. For instance, if the aforementioned factors were recorded, but standard of care or center expertise differed between the controls and treated patients, this may violate positivity. Moreover if standard of care or center expertise differed between the controls and treated patients, this may violate positivity.

**Figure 1**

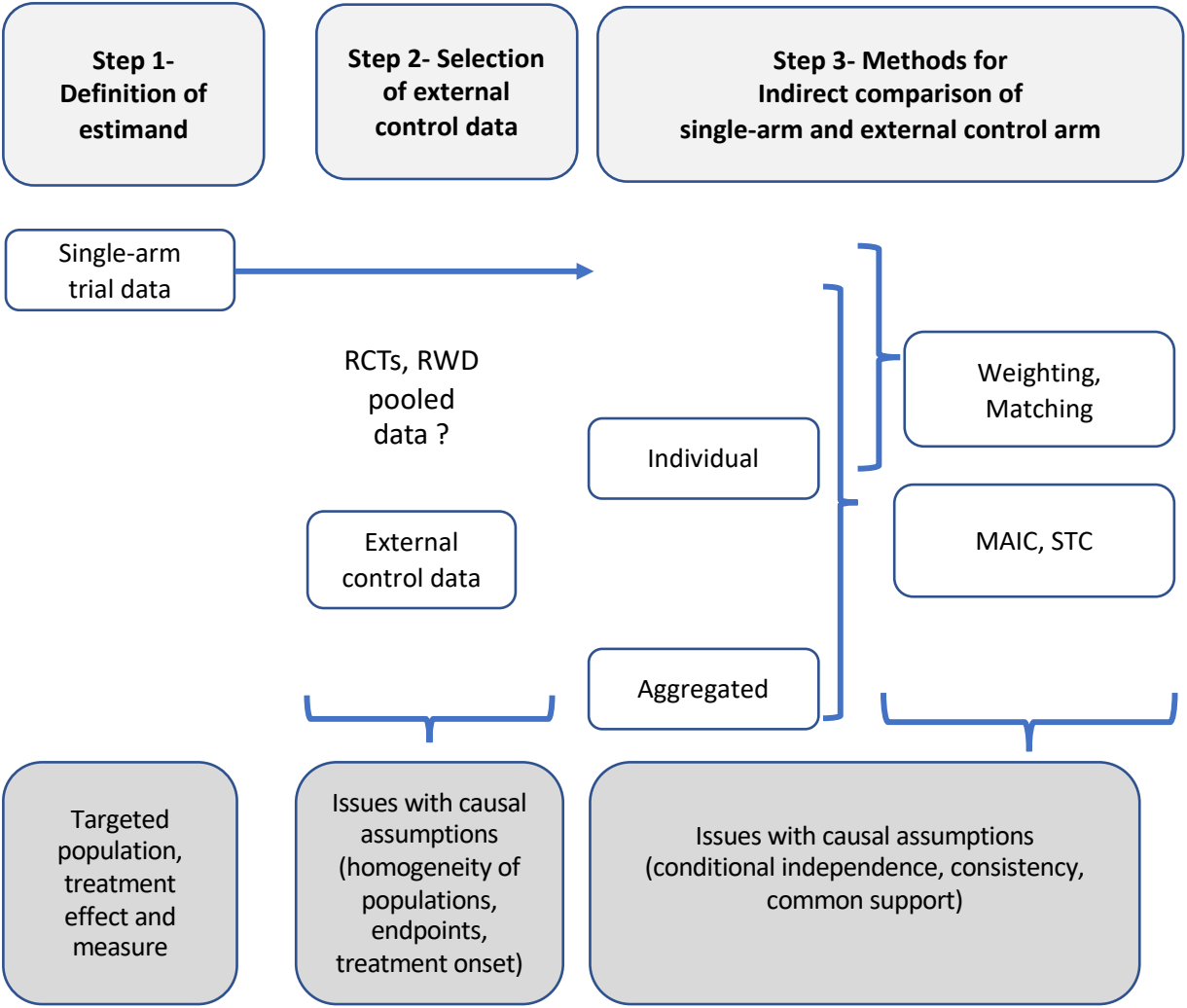

**Figure 2**

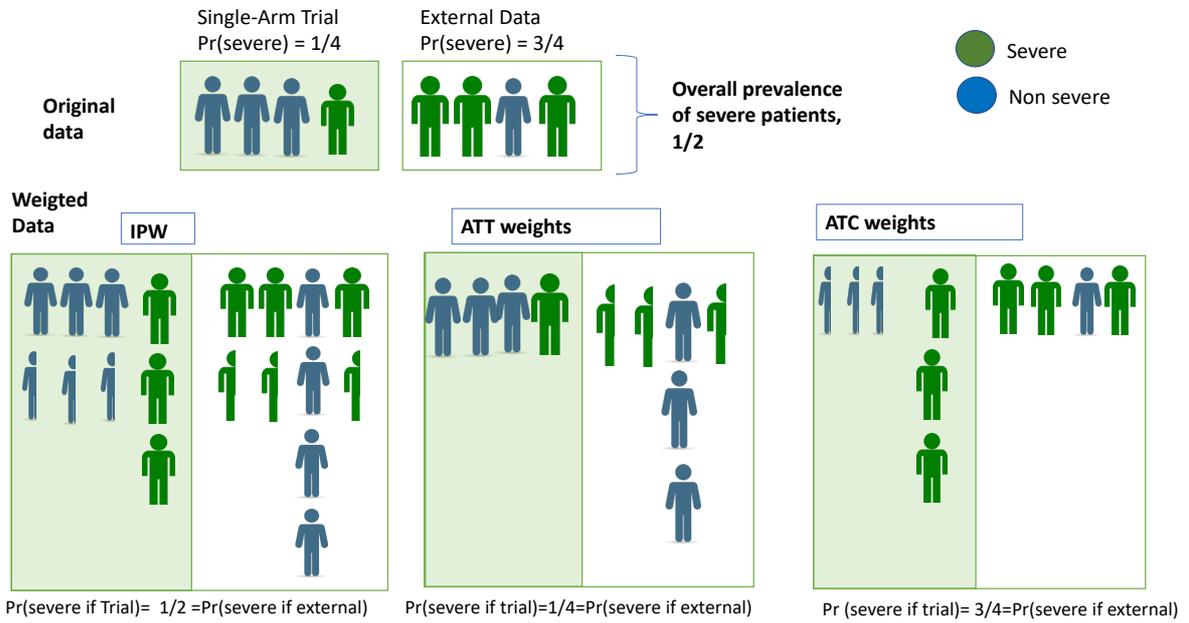